\documentclass[12pt]{article}
\usepackage{epsfig,rotating}
\begin{document}

\begin{titlepage}
\pagenumbering{arabic}
\vspace*{1.5cm}

\begin{center}
{\Large \bf Measurement of the Dalitz plot slope parameters for
\boldmath{$K^- \rightarrow \pi^0\, \pi^0\, \pi^-$} decay using ISTRA+ detector}

\vspace*{1.0cm}
\normalsize
{I.V.~Ajinenko, S.A.~Akimenko, G.I.~Britvich, K.V.~Datsko, A.P.~Filin, 
A.V.~Inyakin, A.S.~Konstantinov, V.F.~Konstantinov, I.Y.~Korolkov, 
V.M.~Leontiev, V.P.~Novikov, V.F.~Obraztsov, V.A.~Polyakov, V.I.~Romanovsky, 
V.I.~Shelikhov, N.E.~Smirnov, O.G.~Tchikilev, V.A.~Uvarov, O.P.~Yushchenko.}
  
\vskip 0.15cm
{\it Institute for High Energy Physics, Protvino, Russia}

\vskip 0.5cm
{V.N.~Bolotov, S.V.~Laptev, A.R.~Pastsjak, A.Yu.~Polyarush, R.Kh.~Sirodeev.}

\vskip 0.15cm
{\it Institute for Nuclear Research, Moscow, Russia}

\end{center}

\vspace*{2.0cm}

\noindent
{\small The Dalitz plot slope parameters $g$, $h$ and $k$ for the 
$K^- \rightarrow \pi^0\,\pi^0\,\pi^-$ decay have been measured using in-flight
decays detected with the ISTRA+ setup operating in the 25~GeV negative 
secondary beam of the U-70 PS. About 252\,K events with four-momenta measured 
for the $\pi^-$ and four involved photons were used for the analysis. The 
values obtained
\begin{center}
$g=0.627\pm0.004(stat)\pm0.010(syst)$,\\
$h=0.046\pm0.004(stat)\pm0.012(syst)$,\\
$k=0.001\pm0.001(stat)\pm0.002(syst)$
\end{center}
are consistent with the world averages dominated by $K^+$ data, but have
significantly smaller errors.}

\end{titlepage}

\newpage
\section{Introduction}

The determination of the Dalitz plot slope parameters for the $K^{\pm} 
\rightarrow (3\pi)^{\pm}$ decays is of interest for the Chiral Perturbation 
Theory (ChPT) and as a check on the direct CP violation. The latter would 
manifest itself by the difference between the $K^+$ and $K^-$ Dalitz plot 
distributions (see for example \cite{Isidori}).

The square of the matrix element of the $\tau^{\,\prime}$ ($K^{\pm} \rightarrow 
\pi^0\pi^0\pi^{\pm}$) decay can be written as
\begin{equation}
\label{eq1}
|A(K^{\pm} \rightarrow 3\pi)|^2 ~ \propto ~ 1+g\,Y+h\,Y^2+k\,X^2+...~,
\end{equation}
where $X = (s_1-s_2)/m_{\pi\,}^2$ and $Y = (s_3-s_0)/m_{\pi\,}^2$ are the 
Dalitz variables, and the parameters $g \div k$ are the ``Dalitz plot slopes''.
Here $s_i = (p_K-p_i)^2$, $s_0 = {1\over3}(s_1+s_2+s_3)$, $p_K$ and $p_i$ are
the $K^{\pm}$ and $\pi_i$ four-momenta ($\pi_3$ is the odd pion).

The direct CP violation could be detected by the observation of the following 
charge asymmetry:
\begin{equation}
\label{eq2}
(\delta g)_{\tau^{\,\prime}} = {{g^+ - g^-} \over {g^+ + g^-}}.
\end{equation}
The theoretical predictions for the asymmetry $(\delta g)_{\tau^{\,\prime}}$ 
in the framework of the Standard Model (SM) were originally spread in the wide
range of $\sim 2\cdot10^{-6} \div 10^{-3}$ \cite{maiani}. Over last years they
have converged to the value of $\sim 10^{-5}$ \cite{Isidori,ambrosimo}. In a 
wide class of possible supersymmetric extensions of the SM larger values are 
possible. For example, in the Weinberg model \cite{weinberg} the value of 
$\sim 2\cdot10^{-4}$ is predicted \cite{shabalin}.

This topic has been attracting significant interest in the recent years. 
The comparison of the results of the latest $K^+$ experiment \cite{batusov} 
with the only one existing $K^-$ measurement \cite{bolotov} leads to the value 
of $(\delta g)_{\tau^{\,\prime}} = 0.117\pm0.022$, i.e. 5.3 sigma effect.

Recently the CP conserving amplitudes for the $K \rightarrow 3\pi$ decays 
were recalculated in ChPT at the next-to-leading order \cite{bijnens}.
In this view a new high statistics measurement for the 
$K \rightarrow 3\pi$ decays is desirable. 

These observations encourage us to perform a new measurement of the Dalitz
plot slope parameters $g$, $h$ and $k$ for the 
$K^- \rightarrow \pi^0\,\pi^0\,\pi^-$ decay, based on the statistics
of about 250\,K events.

\section{Experimental setup}

The experiment has been performed at the IHEP proton synchrotron U-70 with the
experimental apparatus ``ISTRA+'' which is a modification of the ``ISTRA-M'' 
setup \cite{ISTRA-M} and described in some details in our recent papers 
\cite{paper12}, where studies of the $K_{e3}^-$ and $K_{\mu 3}^-$ decays 
have been presented. The setup is located in the negative unseparated 
secondary beam with the following parameters during the measurements: the 
momentum is $\sim25$~GeV/$c$ with $\Delta p/p \sim2\,\%$, the admixture of 
kaons is $\sim3\,\%$, and the total intensity is $\sim3\cdot10^6$ per spill.

\begin{figure}
\begin{turn}{90}
\centering\mbox{\epsfig{file=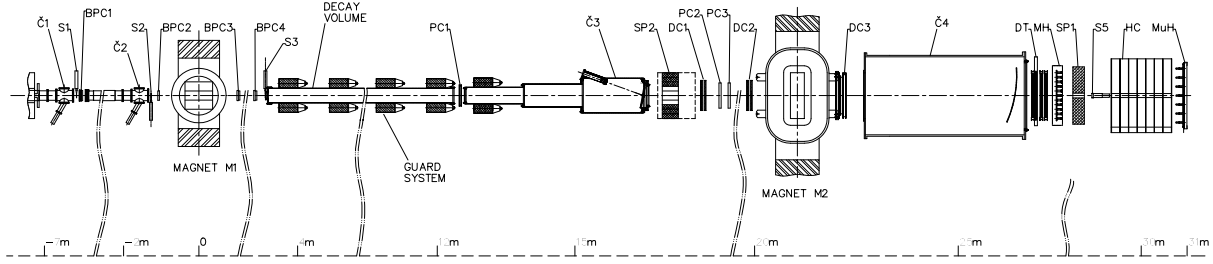,height=\textwidth}}
\end{turn}
\caption{\small The side view of the ``ISTRA+'' detector.}
\label{fig1}
\end{figure}

The side view of the ``ISTRA+'' detector is shown in Fig.\,\ref{fig1}. 
The momentum and the direction of the beam particle, 
deflected by the magnet M1, is measured with 
four proportional chambers BPC1--BPC4. The kaon identification is done by three 
threshold gas Cherenkov counters $\check{\rm C}$0--$\check{\rm C}$2. 
The momenta of the secondary charged particles, deflected in the vertical plane
by the magnet M2, are 
measured with three proportional chambers PC1--PC3, with three drift chambers 
DC1--DC3, and with four planes of drift tubes DT. 
The secondary photons are registered by the lead glass electromagnetic 
calorimeters SP1 and SP2.
To veto low energy photons the decay volume is surrounded by the guard system 
of eight lead glass rings and by the SP2. 
The wide aperture threshold helium Cherenkov counters $\check{\rm C}$3 and
$\check{\rm C}$4 are used to trigger the electrons and are not relevant
for the present paper. 
In Fig.\,\ref{fig1}, HC is a scintillator-iron sampling hadron calorimeter, 
MH is a scintillator hodoscope used to solve the reconstruction ambiguity for 
multitrack events and improve the time resolution of the tracking system, MuH 
is a muon hodoscope.

The trigger is provided by the scintillation counters S1--S5, the Cherenkov 
counters $\check{\rm C}$0--$\check{\rm C}$2, and the sum of the amplitudes 
from the last dinodes of the calorimeter SP1 (see Ref.~\cite{paper12}
for details). The latter serves to suppress the 
$K^- \rightarrow \mu^- \bar{\nu}_{\mu}$ decay. 

\section{Event selection}

About 363\,M and 332\,M events were collected during two physics runs in 
Spring 2001 and Autumn 2001. These experimental data are supported by about 
260\,M events generated with the Monte Carlo program GEANT3 \cite{geant}. 
The Monte Carlo simulation includes a realistic description of the 
experimental setup: the decay volume entrance windows, the track chamber 
windows, gas, sense wires and cathode structures, the Cherenkov counter 
mirrors and gas, the showers development in the electromagnetic calorimeters, 
etc. The details of the reconstruction procedure have been presented in 
Ref.~\cite{paper12}, here only key points relevant to the 
$K^- \rightarrow \pi^-\pi^0\pi^0$ event selection are described.

The data processing starts with the beam particle reconstruction in the 
proportional chambers BPC1--BPC4, and then with the secondary tracks 
reconstruction in the tracking system PC1--PC3, DC1--DC3 and DT. Finally, 
the electromagnetic showers are looked for in the calorimeters SP1 and SP2. 
The method of the photons reconstruction is based on the Monte Carlo
generated patterns of showers. To suppress leptonic decays of kaons 
the particle identification is used \cite{paper12}. The muons are identified 
using the information from the calorimeters SP1 and HC. 
The electrons are identified using the ratio of the energy of the shower, 
detected in the SP1 and associated with the track of the electron, and the 
momentum of the electron.

At the first step of the event selection only the measurements of the beam and 
secondary charged particles are used. Those events are selected which satisfy
the following requirements:
\begin{itemize}
\item[--]
only one beam track and only one negative secondary track are detected;
\item[--]
the probability of the vertex fit, $CL(\chi^2)$, is more than 0.01;
\item[--]
the decay vertex is in the region before the calorimeter SP2
(400 cm $< z <$ 1650 cm), 
and its transverse deviation from the setup axis is less than 10 cm;
\item[--]
the angle between the beam track and the secondary track is more than 2.5 mrad;
\item[--]
the transverse momentum of the secondary track with respect to the beam
direction is less than 150 MeV/$c$;
\item[--]
the secondary track is not identified as an electron or as a muon.
\end{itemize}

\begin{figure}
\centering\mbox{\epsfig{file=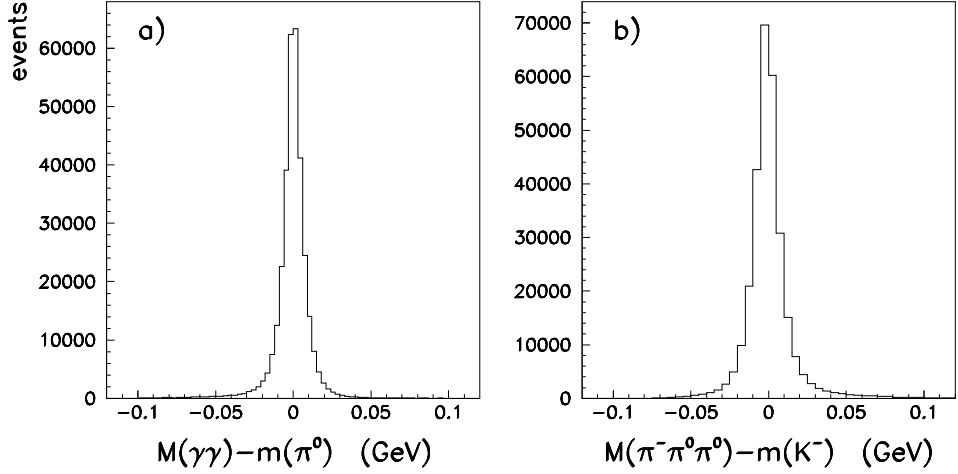,width=0.80\textwidth}}
\caption{\small 
{\bf a)} The deviation $\Delta M(\gamma\gamma) = M(\gamma\gamma)-m(\pi^0)$ of 
the effective mass of the $\gamma\gamma$ pair with the smallest value of 
$|\Delta M(\gamma\gamma)|$ in the selected events with at least four detected 
photons. 
{\bf b)} The deviation $\Delta M(\pi^-\pi^0\pi^0) = M(\pi^-\pi^0\pi^0)-m(K^-)$
of the effective mass of the $\pi^-\pi^0\pi^0$ system with the best photon
combination in the selected events.}
\label{fig2}
\end{figure}

At the second step of the event selection the measurements of the showers in 
the calorimeters SP1 and SP2 are used. The following requirements are used to 
choose the photons (showers):
\begin{itemize}
\item[--]
the distance between the shower in the calorimeter SP1 and the intersection of 
the secondary track with the transverse plane of the SP1 is more than 1.5 cm 
along the magnetic field of the M2 and more than 9 cm in the transverse plane; 
\item[--]
the photon energy is more than 0.7 GeV, but is more than 1.4 GeV
when the photon is detected in three or less cells of the calorimeter SP1;
\item[--]
for events where the secondary track is not associated with any shower in 
the calorimeter SP1 and with any hit in the hodoscope MH the photon energy 
is more than 1.4 GeV;
\item[--]
the energy of the photon found in the multishower cluster is more than 2 GeV.
\end{itemize}
For each selected $\gamma\gamma$ pair the deviation of its effective mass from 
the $\pi^0$ mass, $\Delta M(\gamma\gamma) = M(\gamma\gamma)-m(\pi^0)$, is 
calculated. If this deviation is in the range of $|\Delta M(\gamma\gamma)|<50$ 
MeV, the $\gamma\gamma$ pair is considered as a candidate of the $\pi^0$ decay.
Then, if this $\gamma\gamma$ pair is taken as a $\pi^0$ decay, the four-momenta 
of its photons are multiplied by a factor $\lambda=m(\pi^0)/M(\gamma\gamma)$.
The $\pi^0$ detection is illustrated in Fig.\,\ref{fig2}a, where the spectrum 
of the effective masses of the ``best'' $\gamma\gamma$ pairs, i.e. the pairs 
with the smallest value of $|\Delta M(\gamma\gamma)|$, is shown for the 
selected events with at least four detected photons. In further selections all 
combinatorial $\pi^0$ candidates are used, not only the ``best''.

At the third step of the event selection the sample of the 
$K^- \rightarrow \pi^-\gamma\gamma\gamma\gamma$ events with two 
$\pi^0 \rightarrow \gamma\gamma$ decays is collected.
For this sample the further selection is done by the requirements 
that the measured value of the kaon mass is in the range of 
$|M(\pi^-\pi^0\pi^0)-m(K^-)| < 80$ MeV (see Fig.\,\ref{fig2}b) and then the 
event passes the kinematical 6C-fit for the $K^- \rightarrow \pi^-\pi^0\pi^0$ 
hypothesis. The efficiency of the last cut is about 89\%. Both requirements 
are considered for all possible combinations of photons and the best 6C-fit 
hypothesis is chosen. 

Using the mentioned above selection criteria for the 
$K^- \rightarrow \pi^-\pi^0\pi^0$ decay we have collected 252\,K 
completely reconstructed events. 
The corresponding numbers of accepted Monte Carlo events are about ten times 
larger than in the experiment. 
The surviving background, which is mainly due to kaon decays to other allowed 
modes, is estimated from the Monte Carlo simulation to be less than 0.04\%.

The detailed event reduction statistics is given in Table~\ref{tab1}. 
The difference in the fraction of the selected events for two runs is due
to the higher threshold in the SP1 at the trigger level for the second run,
i.e. the initial statistics of the first run contains more
$K^- \rightarrow \mu^- \bar{\nu}_{\mu}$ decays than the initial statistics of 
the second run.
\begin{table}[bth]
\caption{The event reduction statistics for two physics runs.}
\renewcommand{\arraystretch}{1.5}
\begin{center}
\begin{tabular}{|l|r|r|}
\hline
Run&{\it Spring~2001~}&{\it Autumn~2001~}\\  
\hline \hline
Total number of events& 363\,M & 332\,M \\
\hline
Beam track reconstructed& 269\,M & 248\,M \\
\hline
Secondary track(s) reconstructed& 121\,M & 124\,M \\                    
\hline
Number of events written on DST& 93\,M & 108\,M \\
\hline \hline
$K^-$ and $\pi^-$ selected (all cuts of the 1st step)& 4882\,K & 7432\,K \\
\hline
at least four $\gamma$ selected (all cuts of the 2nd step)& 83\,K & 244\,K \\
\hline
two $\pi^0$ selected&
 69\,K & 218\,K \\
\hline
$K^- \rightarrow \pi^-\pi^0\pi^0$ selected (all cuts of the 3rd step)&
 59\,K & 193\,K \\
\hline
\end{tabular}
\end{center}
\label{tab1}
\end{table}
\clearpage

\section{Analysis}

\begin{figure}[h]
\centering\mbox{\epsfig{file=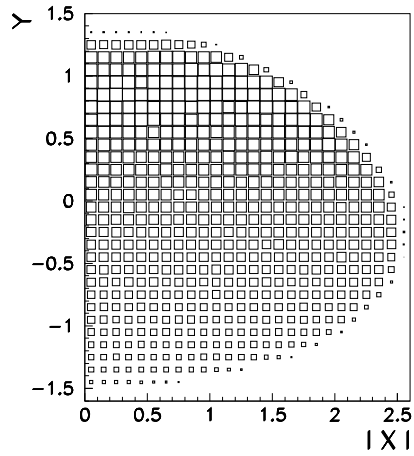,width=0.3505\textwidth}}
\caption{\small The Dalitz plot, 
$Y = (s_3-s_0)/m_{\pi\,}^2$ {\it versus} $X = (s_1-s_2)/m_{\pi\,}^2$, for the 
selected $K^- \rightarrow \pi^-\pi^0\pi^0$ events (uncorrected).}
\label{fig3}
\end{figure}

\begin{figure}[h]
\centering\mbox{\epsfig{file=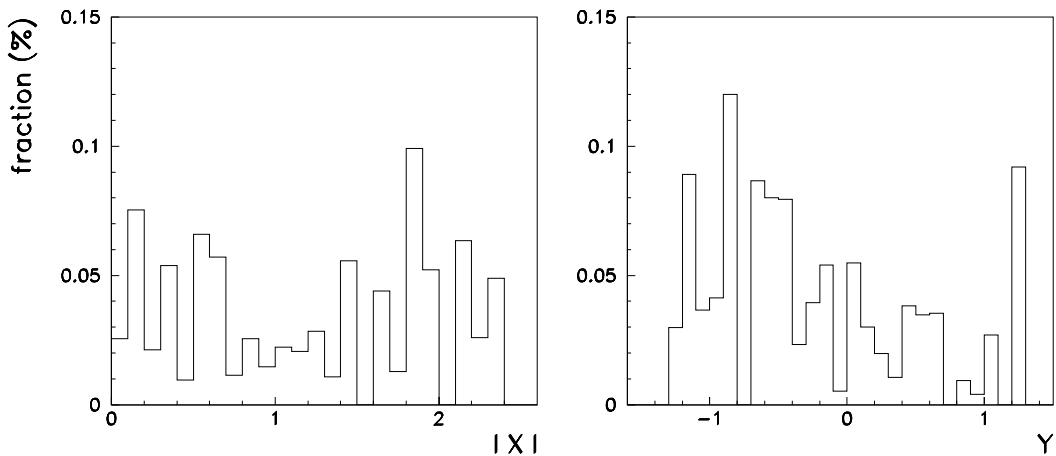,width=0.885\textwidth}}
\caption{\small The fraction of the background contamination as a function of 
the Dalitz plot variable estimated from Monte Carlo simulation.}
\label{fig4}
\end{figure}

To determine the parameters $g$, $h$ and $k$ in Eq.\,(\ref{eq1}) the 
uncorrected distribution $\rho(X,Y)$ of the event density on the Dalitz plot 
was analysed. This distribution is shown in Fig.\,\ref{fig3} for the selected 
$K^- \rightarrow \pi^-\pi^0\pi^0$ events.
At first the background contamination was subtracted from the Dalitz plot.
The contamination was estimated from the Monte Carlo simulation of the particle
interaction with the material of the detector and of the kaon decay including
all decay modes with the branching ratios more than 1\%. The corresponding 
branching ratios and matrix elements in this simulation were taken from the 
PDG \cite{PDG}.
In Fig.\,\ref{fig4} the fraction of the background contamination is shown as 
a function of the Dalitz plot variables $|X|$ and $Y$.

\begin{figure}[h]
\centering\mbox{\epsfig{file=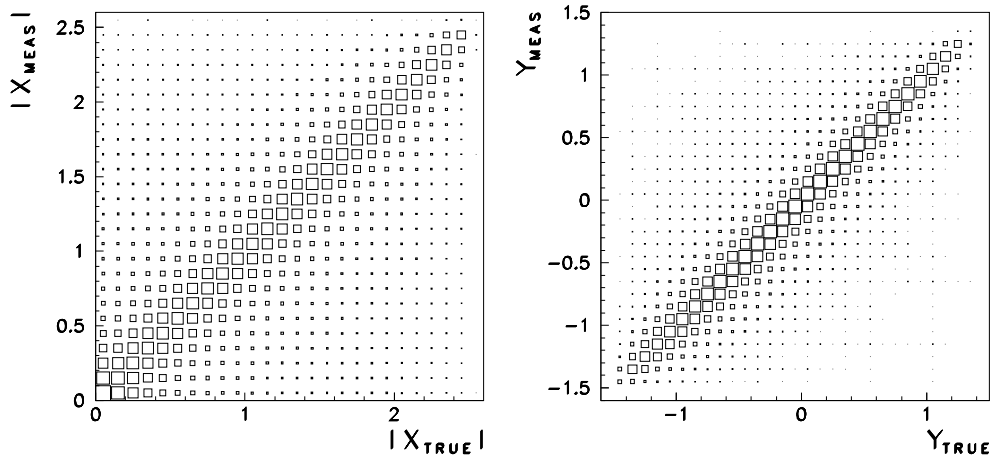,width=0.845\textwidth}}
\caption{\small The ``measured'' values of the Dalitz plot variables $|X|$ and 
$Y$ versus the ``true'' ones estimated from Monte Carlo simulation.}
\label{fig5}
\end{figure}

The background subtracted distribution $\rho^{\,\prime}(X,Y)$ was fitted by the 
method of least squares with the function: 
\begin{equation}
\label{eq3}
\rho^{\,\prime}(X,Y) ~ \propto ~
F_1 (X,Y) + g \, F_2 (X,Y) + h \, F_3 (X,Y) + k \, F_4 (X,Y),
\end{equation}
where $F_k (X,Y)$ are the distributions for the $w_k$--weighted Monte Carlo 
$K^- \rightarrow \pi^-\pi^0\pi^0$ events generated with the constant matrix 
element and reconstructed with the same program as for the real events.
The weight factors $w_1=1$, $w_2=Y_{true}$, $w_3=Y_{true}^2$ and 
$w_4=X_{true}^2$ are given by the ``true'' values of $X$ and $Y$, but the bins
of $F_k (X,Y)$ are given by the ``measured'' ones.
This method allows to avoid the systematic errors \cite{anikeev} due to the 
``migration'' of the events on the Dalitz plot because of the finite 
experimental resolution.
Fig.\,\ref{fig5} illustrates the Monte Carlo estimated experimental resolution 
for the variables $|X|$ and $Y$, where the ``measured'' values are shown versus
the ``true'' ones.

\section{Results}

\begin{figure}
\centering\mbox{\epsfig{file=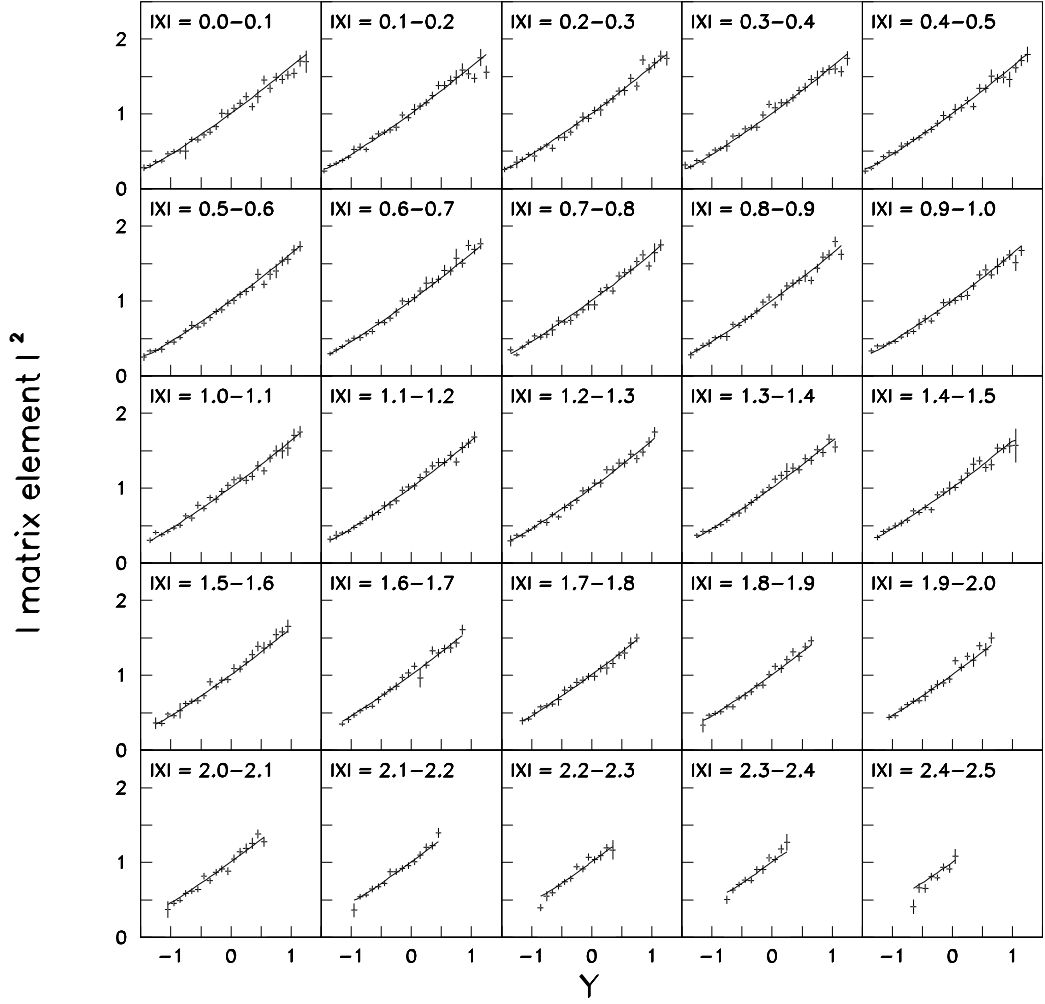,width=0.87\textwidth}}
\caption{\small
The matrix element dependence on the variable $Y$ in the different intervals 
of the variable $|X|$ for the $K^- \rightarrow \pi^-\pi^0\pi^0$ decay. 
The curves are the result of the fit to the function (\ref{eq3}).}
\label{fig6}
\end{figure}

\begin{figure}
\centering\mbox{\epsfig{file=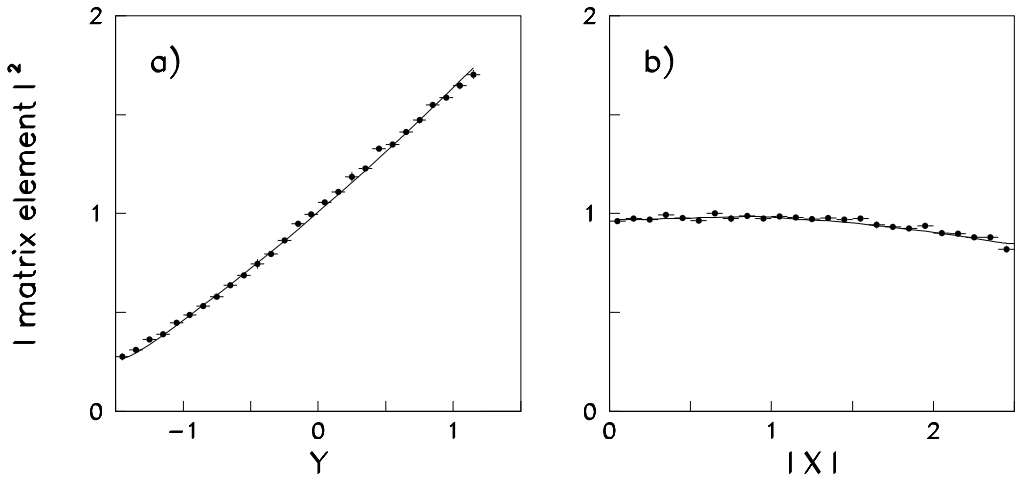,width=0.86\textwidth}}
\caption{\small
The integrated dependences of the matrix element on the variables {\bf a)} $Y$ 
and {\bf b)} $|X|$ for the $K^- \rightarrow \pi^-\pi^0\pi^0$ decay. 
The curves are the result of the fit to the function (\ref{eq3}).}
\label{fig7}
\end{figure}

The result of the least squares fit is illustrated in Fig.\,\ref{fig6}, where 
the matrix element 
\begin{equation}
\label{eq4}
|A(K^- \rightarrow \pi^-\pi^0\pi^0)|^2 ~ = ~
C \cdot {{\rho^{\,\prime}(X,Y)} \over {F_1 (X,Y)}}
\end{equation}
for the $K^- \rightarrow \pi^-\pi^0\pi^0$ decay is shown as a function 
of $Y$ in the different intervals of $|X|$.
The normalization constant $C$ in Eq.\,(\ref{eq4}) provides the value of 
$|A|^2 = 1$ at the point ($X$=0, $Y$=0). The integrated dependences of the 
matrix element on the variables $Y$ and $|X|$ are shown in Fig.\,\ref{fig7}.

The values of the Dalitz plot slope parameters are found to be
\begin{center}
$g ~=~ 0.627 \pm 0.004 \pm 0.010,$\\
$h ~=~ 0.046 \pm 0.004 \pm 0.012,$\\
$k ~=~ 0.001 \pm 0.001 \pm 0.002~$\\
\end{center}
with $\chi^2/ndf = 502/558$.
Here the first errors are statistical and the second ones are systematic.

In the determination of the systematic errors of the slope parameters
measurement the following contributions were taken into account.
\begin{itemize}
\item[--]
Two samples of the events collected in the runs (Spring and Autumn 2001)
with some differences in characteristics of the setup were analysed separately 
(the corresponding contributions to the systematic errors are 
$\Delta g$ = 0.001, $\Delta h$ = 0.001 and $\Delta k$ = 0.0002).
\item[--]
To avoid some uncertainties at the edge of the Dalitz plot the extreme bins 
of this plot were cut ($\Delta g$ = 0.001, $\Delta h$ = 0.002,
$\Delta k$ = 0.0004).
\item[--]
The energy threshold of the selected photons was increased from the value of
0.7 GeV to 2 GeV ($\Delta g$ = 0.006, $\Delta h$ = 0.007, $\Delta k$ = 0.001).
\item[--]
The mass and energy ranges used in the event selection criteria were varied 
from the value of 30 MeV to 80 MeV with and without introducing a factor
$\lambda=m(\pi^0)/M(\gamma\gamma)$ ($\Delta g$ = 0.003, $\Delta h$ = 0.004,
$\Delta k$ = 0.001).
\item[--]
The background contamination estimated from the Monte Carlo simulation
was not subtracted from the Dalitz plot before the least squares fit
($\Delta g$ = 0.001, $\Delta h$ = 0.001, $\Delta k$ = 0.0002).
\item[--]
The particle identification of the secondary track was not used 
($\Delta g$ = 0.001, $\Delta h$ = 0.001, $\Delta k$ = 0.0003).
\item[--]
The electromagnetic showers in the calorimeter SP2 were not used in the photon 
reconstruction ($\Delta g$ = 0.006, $\Delta h$ = 0.007, $\Delta k$ = 0.0014).
\item[--]
The upper edge of the decay vertex position was varied along the setup axis
between the chamber PC1 and the calorimeter SP2 ($\Delta g$ = 0.004, 
$\Delta h$ = 0.005, $\Delta k$ = 0.001).
\item[--]
The cut applied to the angle between the beam and secondary tracks was varied 
from 1.5 to 3.5 mrad ($\Delta g$ = 0.001, $\Delta h$ = 0.001,
$\Delta k$ = 0.0003).
\end{itemize}

\section{Summary and conclusion}

\begin{figure}
\centering\mbox{\epsfig{file=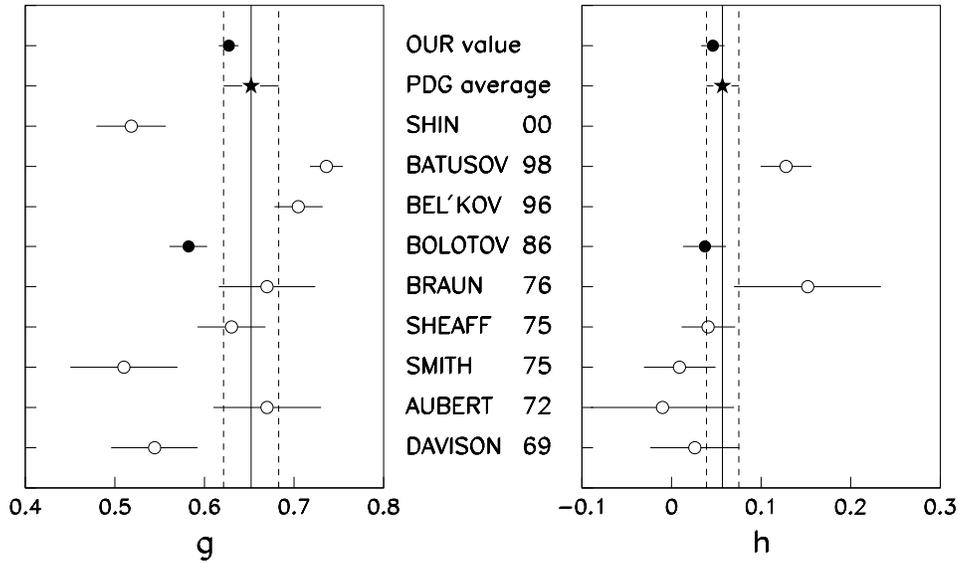,width=0.8\textwidth}}
\caption{\small
The Dalitz plot slope parameters $g$ and $h$ for the 
$K^- \rightarrow \pi^-\pi^0\pi^0$ (solid circles),
$K^+ \rightarrow \pi^+\pi^0\pi^0$ (open circles) and
$K^{\pm} \rightarrow \pi^{\pm}\pi^0\pi^0$ (solid stars) decays.}
\label{fig8}
\end{figure}

The Dalitz plot slope parameters for the $K^- \rightarrow \pi^-\pi^0\pi^0$ 
decay have been measured using the ``ISTRA+'' spectrometer.
It is an update of our preliminary presentation \cite{paper3}.
The results of our measurement, the world averages \cite{PDG} and the results
of previous experiments [6, 7, 15 -- 21] on the 
$K^{\pm} \rightarrow \pi^{\pm}\pi^0\pi^0$ decays are presented in 
Fig.\,\ref{fig8}.
Among the previous experiments there are eight measurements of the $K^+$ decay, 
but only one of the $K^-$ decay. 
Our values of the slope parameters $g$ and $h$ are consistent with the world 
averages dominated by $K^+$ measurements.
The difference between the values of the linear slope $g$ obtained for
the $K^- \rightarrow \pi^-\pi^0\pi^0$ decay in our experiment and in another 
one \cite{bolotov} is 1.9 standard deviations.

Using the same rules as in the PDG \cite{PDG} (without the data 
\cite{belkov,shin} obtained from the linear fit only) and adding
our measurement, one can obtain 
the world average values of the linear slope $g$ for the $K^+$ and $K^-$
decays separately: $g^+ = 0.684 \pm 0.033$ and $g^- = 0.617 \pm 0.018$.
They give for the charge asymmetry (\ref{eq2}) the value of
$(\delta g)_{\tau^{\,\prime}} = 0.051 \pm 0.028$.

The analytical expressions of the CP conserving $K \rightarrow 3\pi$ amplitudes
with single parameter functions, recalculated in the framework of ChPT at the
next-to-leading order, were fitted in Ref. \cite{bijnens} to all available 
$K \rightarrow 2\pi$ and $K \rightarrow 3\pi$ data.
The result of this global fit gives the following predictions for physical 
observables of the $K^{\pm} \rightarrow \pi^{\pm} \pi^0 \pi^0$ decays:
$g$ = 0.638, $h$ = 0.074 and $k$ = 0.0045.
These predictions agree within $1 \div 2$ standard deviations with our results.

\smallskip
{\it The work is supported by the RFBR contract no. 03-02-16330.}

\end{document}